\documentclass[prl,twocolumn,showkeys,superscriptaddress,amsmath,amsfonts]{revtex4-1}
\usepackage[latin9]{inputenc}
\usepackage{amstext}
\usepackage{graphicx}
\usepackage{esint}
\usepackage{cancel}

\makeatletter
\usepackage{dcolumn}
\usepackage{bm}
\usepackage{times}
\usepackage{color}

\makeatother

\begin{document}

\title{Defect-Induced Orbital Polarization and Collapse of Orbital Order in
  Doped Vanadium Perovskites }

\author{Adolfo Avella}

\affiliation{Dipartimento di Fisica ``E.R. Caianiello'', Universit\`a degli
Studi di Salerno, I-84084 Fisciano (SA), Italy}
\affiliation{CNR-SPIN, UOS di Salerno, I-84084 Fisciano (SA), Italy}
\affiliation{Unit\`a CNISM di Salerno, Universit\`a degli Studi di Salerno,
I-84084 Fisciano (SA), Italy}

\author{Andrzej M. Ole\'s}

\affiliation{\mbox{Marian Smoluchowski Institute of Physics, Jagiellonian
University, Prof. S. \L{}ojasiewicza 11, PL-30348 Kraków, Poland}}
\affiliation{Max-Planck-Institut für Festkörperforschung,
Heisenbergstrasse 1, D-70569 Stuttgart, Germany}

\author{Peter Horsch}

\affiliation{Max-Planck-Institut für Festkörperforschung,
Heisenbergstrasse 1, D-70569 Stuttgart, Germany}

\date{21 January 2019}

\begin{abstract}
We explore mechanisms of orbital-order decay in the doped Mott 
insulators $R_{1-x}$(Sr,Ca)$_x$VO$_3$ \mbox{($R=\,$Pr,Y,La)} caused by 
charged (Sr,Ca) defects. Our unrestricted Hartree-Fock analysis focuses 
on the combined effect of random charged impurities and associated 
doped holes up to $x=0.5$. The study is based on a generalized 
multi-band Hubbard model for the relevant vanadium $t_{2g}$ electrons, 
and includes the long-range (i) Coulomb potentials of defects and 
(ii) electron-electron interactions. We show that the rotation of 
$t_{2g}$ orbitals, induced by the electric field of defects, is a very
efficient perturbation that largely controls the suppression of
orbital order in these compounds. We investigate the inverse
participation number spectra and find that electron states remain
localized on few sites even in the regime where orbital order is
collapsed. From the change of kinetic and superexchange energy we can
conclude that the motion of doped holes, which is the dominant effect
for the reduction of magnetic order in high-$T_c$ compounds,
is of secondary importance here.
\end{abstract}

\maketitle

%%%%%%%%%%%%%%%%%%%%%%%%%%%%%%%%%%%%%%%%%%%%%%%%%%%%%%%%%%%%%%%%%%%%%%%%
%    INTRODUCTION
%%%%%%%%%%%%%%%%%%%%%%%%%%%%%%%%%%%%%%%%%%%%%%%%%%%%%%%%%%%%%%%%%%%%%%%%

Doping of Mott insulators is a central topic in materials science
\cite{Kei15,Ima98}, cold gases \cite{Gru18}, and many-body theory
\cite{Zho17} --- firstly because of the intriguing  origin of the
insulating state, due to strong electron correlations, and secondly
owing to the amazing features that can emerge when they are doped,
such as superconductivity in cuprates
\cite{Uch91,Kas98,Lee06,Sca12,Tacon,Fra15}, magneto- and thermo-electric
effects in manganites \cite{Tok06,Dag01,Qui98,Kil99,Gio08}
and heterostructures \cite{Yun07,Cha12,Cao16,Kue17}. Yet, often such 
systems remain insulating when doped, although transitions
into metallic or superconducting states were expected \cite{Lei14}.
The cubic vanadium perovskites show, despite strong quantum orbital
fluctuations \cite{Kha01,Kha05,Ray07,Yan04,Reu12}, an unusual
gradual decay of orbital and spin order and a not-well-defined
crossover into a poor metallic state at high doping $x$, e.g.,
$x=0.18$ in La$_{1-x}$Sr$_x$VO$_3$ and $x=0.50$ in
Y$_{1-x}$Ca$_x$VO$_3$ \cite{Kas93,Tok00,Fuj05,Fuj08}. This makes them 
an ideal platform for the study of charged defects and of their 
interaction with doped holes in systems with spin-orbital degrees
of \mbox{freedom \cite{Ave15,Ave18}}.

Vanadates are Mott insulators where the $t_{2g}$ electrons form a
$d^2$ configuration with a $S=1$ spin at each V ion. A small crystal
field (CF) lowers the energy of $xy$ orbitals by $\Delta_c\simeq 0.1$
eV with respect to $\{yz,zx\}$ orbital doublet
\cite{Tok00,Fuj05,Fuj08,Ren00,Miy06,Sol06,Fuj10,Yan11},
which is the source of strong orbital quantum fluctuations
\cite{Kha01,Kha05,Ray07,Yan04,Reu12}.
The breaking of an almost perfect cubic crystal symmetry leads to
highly anisotropic electronic states. The undoped systems reveal two
distinct spin-orbital ordered ground states. In systems with a large
$R$-ion radius, as LaVO$_3$, the ground state has coexisting spin
$C$-type AF ($C$-AF) and \mbox{$G$-type} alternating orbital ($G$-AO) 
order \cite{Ren00,Miy06,Sol06,Fuj10,Yan11}, which is stabilized by the
effective spin-orbital superexchange interactions \cite{Kha01,Hor08}.
A second type of complementary \mbox{$G$-AF/$C$-AO} spin-orbital order 
results
from a competition of superexchange and Jahn-Teller (JT) interactions
\cite{Kha01} and occurs in undoped $R$VO$_3$ perovskites with small
radii of $R$ ions, as in YVO$_3$ \cite{Ren00,Miy06,Sol06,Fuj10,Yan11}.

Motivations to analyze the role of charged defects are:
(i)~the surprising discovery that the $G$-AF/$C$-AO ground state 
of YVO$_3$ changes already at $x\simeq 1\%$ Ca doping into the
\mbox{$C$-AF/$G$-AO} state \cite{Ren00,Nog00,Bla01}, and
(ii) the stability of the latter phase up to high doping
\cite{Fuj08,Sag08,Ree16}. The fragility of \mbox{$G$-AF/$C$-AO} order 
relative to $C$-AF/$G$-AO phase was explained by a double exchange
process for the doped hole bound to the charged defect, triggered by 
the FM correlations in the \mbox{$C$-AF} state \cite{Hor11}.
Subsequent studies have shown that the holes in the $C$-AF/$G$-AO state 
are confined and bound to the charged defects, leading to a gradual 
decay of order proportional to doping, yet not to its 
\mbox{collapse \cite{Ave18}}.

\begin{figure}[t!]
\includegraphics[width=.85\columnwidth]{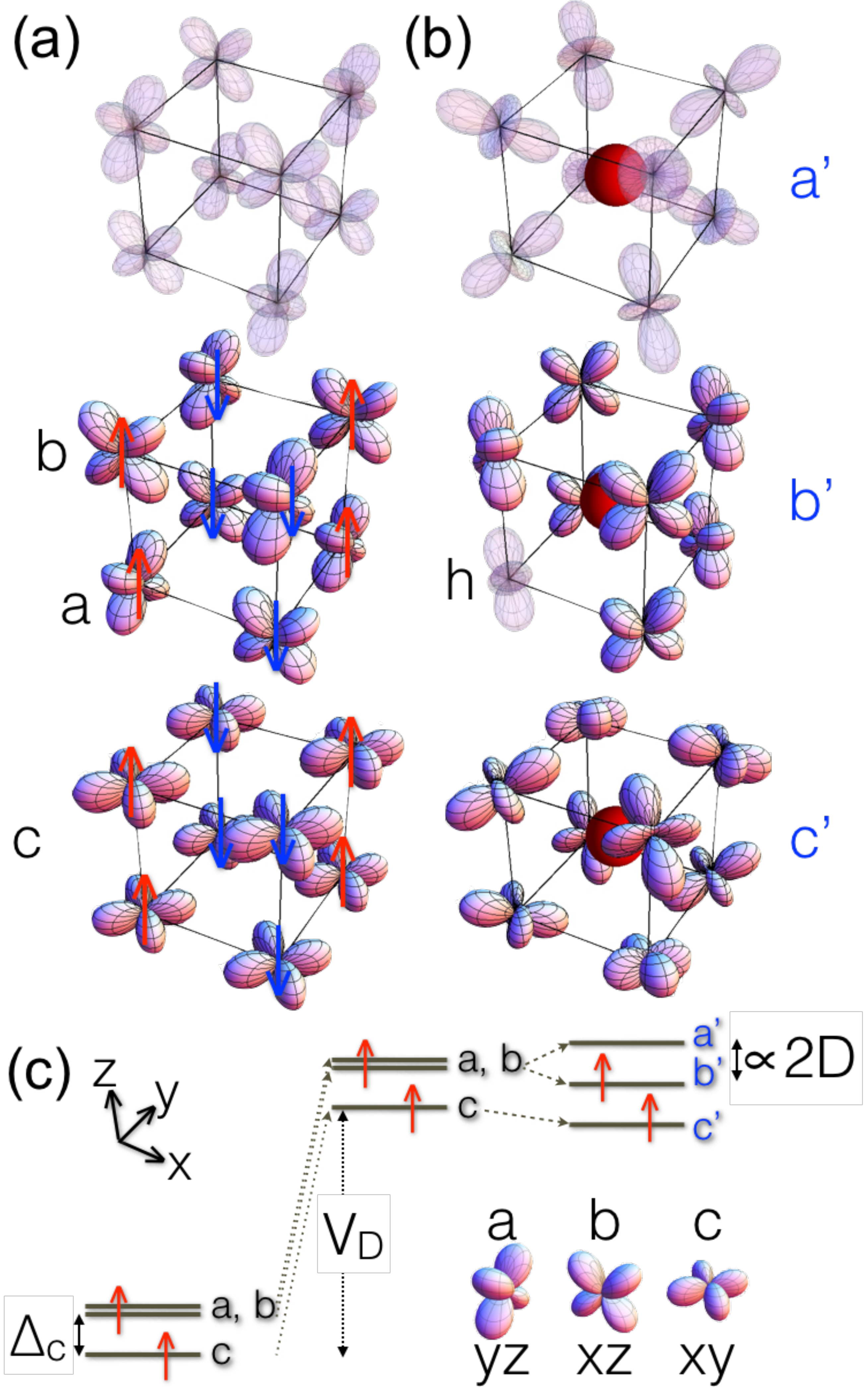}
\protect\caption{
Schematic view of occupied and unoccupied (grayed out) $t_{2g}$
V orbitals for:
(a) $G$-AO order in undoped $R$VO$_3$ with \mbox{$C$-AF} spin order
marked by red/blue arrows, and
(b) a \textit{defect cube} around a Ca$^{2+}$ defect (red sphere)
in $R_{1-x}$Ca$_x$VO$_3$, with $\{a',b',c'\}$ orbitals in the large
$\mathcal{D}$ limit. Finite $\mathcal{D}$ modifies the standard 
$t_{2g}$ basis $\{a,b,c\}$ at each V site to $\{a',b',c'\}$; the lowest 
orbitals $\{c',b'\}$ are occupied at all but the hole (h) site.
(c) $t_{2g}$ orbital energies at a V ion for $\mathcal{D}\sim\Delta_c/2$, 
with the $\{a',b'\}$ doublet split by $2\mathcal{D}$.
}
\label{fig:orbs}
\end{figure}

In this Letter, we investigate the doping dependence of the orbital 
order (OO) in doped vanadates and explain its collapse. We
find that the dominant decay mechanism is the rotation of $t_{2g}$
electron states induced by the Coulomb potential of defects. This
orbital polarization involves all $t_{2g}$ orbitals at V ions
surrounding the defect \cite{Ave13}, i.e., on the
\textit{defect cube}, see Fig. \ref{fig:orbs}. Interestingly, the OO
collapse is visible in the moderate delocalization of the states in
the upper Hubbard band (UHB) and identified as $d^2\rightarrow d^3$
high spin transitions at V ions on the defect cubes.

The Hamiltonian for the $t_{2g}$ electrons in
$R_{1-x}$Ca$_{x}$VO$_{3}$,
\begin{equation}
{\cal H}_{t2g}\!={\cal H}_{\rm Hub}+{\cal H}_{{\rm pol}}
+\sum_{i<j}v(r_{ij})\hat{n}_i\hat{n}_j
+\sum_{mi}v(r_{mi})\hat{n}_i,
\label{Ht2g}
\end{equation}
includes the extended degenerate Hubbard model 
${\cal H}_{\rm Hub}$ \cite{Ole07}, orbital-polarization term 
${\cal H}_{{\rm pol}}$ \cite{Ave13}, and two last terms stand for 
$t_{2g}$ electron-electron interactions and the repulsive potential
of Ca defects. Both are determined by the Coulomb interaction
$\propto v(r)\equiv{e^2}/{\varepsilon_c r}$, where
$\varepsilon_c\simeq 5$ \cite{Hor11} is the dielectric constant of the
core electrons, and $r$ is the distance between interacting charges of:
(i)~two V ions at sites $i$ and $j$ with
\mbox{$r_{ij}=|\mathbf{r}_i-\mathbf{r}_j|$}, and
(ii)~(Ca,Sr) defect at site $m$ and a $t_{2g}$ electron at a V ion at 
site $i$, with \mbox{$r_{mi}=|\mathbf{R}_m-\mathbf{r}_i|$}.
We emphasize that the latter term acts as a potential from all 
defects on the $t_{2g}$ electron charge
$\hat{n}_i=\sum_{\alpha\sigma}\hat{n}_{i\alpha\sigma}$,
with \mbox{$\hat{n}_{i\alpha\sigma}\!
=\hat{d}_{i\alpha\sigma}^{\dagger}\hat{d}_{i\alpha\sigma}^{}$}.

The hopping of the $t_{2g}$ electrons $\propto t\equiv(dd\pi)$ in
${\cal H}_{\rm Hub}$ is two-dimensional and orbital flavor conserving
\cite{Hor11,Ave13,Ole07,Dag11}, which has peculiar consequences
for hole propagation \cite{Ish05,Dag08,Bis15,Bie16,Yam18}.
Below we denote the $t_{2g}$ orbitals $\{yz,xz,xy\}$ by the cubic
directions $\{a,b,c\}$, respectively, for which the hopping is 
forbidden \cite{Kha00} (see Fig.~\ref{fig:orbs}). Intraatomic Coulomb
interactions are parametrized by intraorbital $U$ and Hund's exchange
$J_H$. The rotational invariant form \cite{Ole83} is essential for
multi-orbital models when orbitals and/or spins rotate
\cite{Ave13,Ant12}. The cubic symmetry of the spin-orbital structure
is broken by a CF term $\propto\Delta_c$, which favors the $c^1(a/b)^1$
electronic configuration at V$^{3+}$ ions. The 2nd electron can select
between two degenerate orbitals $\{a,b\}$, according to the
spin-orbital superexchange interaction that emerges from the present
Hubbard model \cite{Kha01}, see Fig. \ref{fig:orbs}(a).

A~Ca$^{2+}$ defect in the lattice of Y$^{3+}$ ions in
Y$_{1-x}$Ca$_x$VO$_3$ acts effectively as a negative charge, which
repels all vanadium electrons on a defect cube by $V_{\rm D}\equiv v(d)$,
as shown in Fig.~\ref{fig:orbs}(c). As we are dealing with a Mott
insulator the upward shift creates defect states in the Mott-Hubbard 
gap \cite{Hor11}. In this work, we focus on another effect of the 
defect's charge that is displayed in Fig.~\ref{fig:orbs}(b).
The $t_{2g}$ vanadium orbitals on a defect cube rotate
to reduce their Coulomb energy in the electric field of the defect.
This rotation is described by \cite{Ave13},
\begin{equation}
\mathcal{H}_{\rm pol}=
\mathcal{D}\!\sum_{m,i\in\mathcal{C}_m \atop \alpha\neq\beta,\sigma}\!\!
\lambda_{\alpha\beta}(\mathbf{r}_i\!-\!\mathbf{R}_m)
\left(\hat{d}_{i\alpha\sigma}^{\dag}\hat{d}_{i\beta \sigma}^{}+
      \hat{d}_{i\beta \sigma}^{\dag}\hat{d}_{i\alpha\sigma}^{}\right)\!.
\label{pol}
\end{equation}
The orbital-polarization parameter $\mathcal{D}$ is defined by the
matrix element
\mbox{$\langle i\alpha|v(|\mathbf{r}_i\!-\!\mathbf{R}_m|)|i\beta\rangle
\equiv\mathcal{D}\lambda_{ab}\left(\mathbf{r}_i\!-\!\mathbf{R}_m\right)$}.
Here, we shall treat $\mathcal{D}$ as a free parameter.
The sign of the matrix element is encoded in
$\lambda_{\alpha\beta}(\mathbf{r}_i\!-\!\mathbf{R}_m)\!=\!\pm 1$
and depends on the vector $\mathbf{r}_i\!-\!\mathbf{R}_m$.
For the $\{a,b\}$ doublet we have \cite{Hor11},
\begin{eqnarray}
\lambda_{ab}\left(\mathbf{r}_{i}\!-\!\mathbf{R}_{m}\right) &
= & \left\{
\begin{array}{ccc}
1 & \text{if} &
(\mathbf{r}_i\!-\!\mathbf{R}_m)\parallel(111),(11\bar{1}),
\nonumber \\
-1 &\text{if} &
(\mathbf{r}_i\!-\!\mathbf{R}_m)\parallel(\bar{1}11),(1\bar{1}1).
\end{array}\right.
\end{eqnarray}
Signs of all other
$\lambda_{\alpha\beta}$ are obtained by cubic symmetry,
see the Supplemental Material \cite{suppl}.

The effect of orbital polarization (\ref{pol}) on vanadium ions around 
a Ca defect is shown in Fig. \ref{fig:orbs}(b) for the large 
$\mathcal{D}$ case. The actual form of the rotated $\{a',b',c'\}$ 
orbitals depends on the corner of the defect cube under analysis. 
The orbitals are here classified according to their energy, see Fig. 
\ref{fig:orbs}(c). This perturbation of the $G$-type OO is expected to 
be a strong effect as it involves the orbitals of all eight V ions in 
a defect cube. It competes with the CF, JT and the superexchange terms,
which stabilize the $C$-AF/$G$-AO order in LaVO$_3$.

\begin{figure}[b!]
\includegraphics[width=\columnwidth]{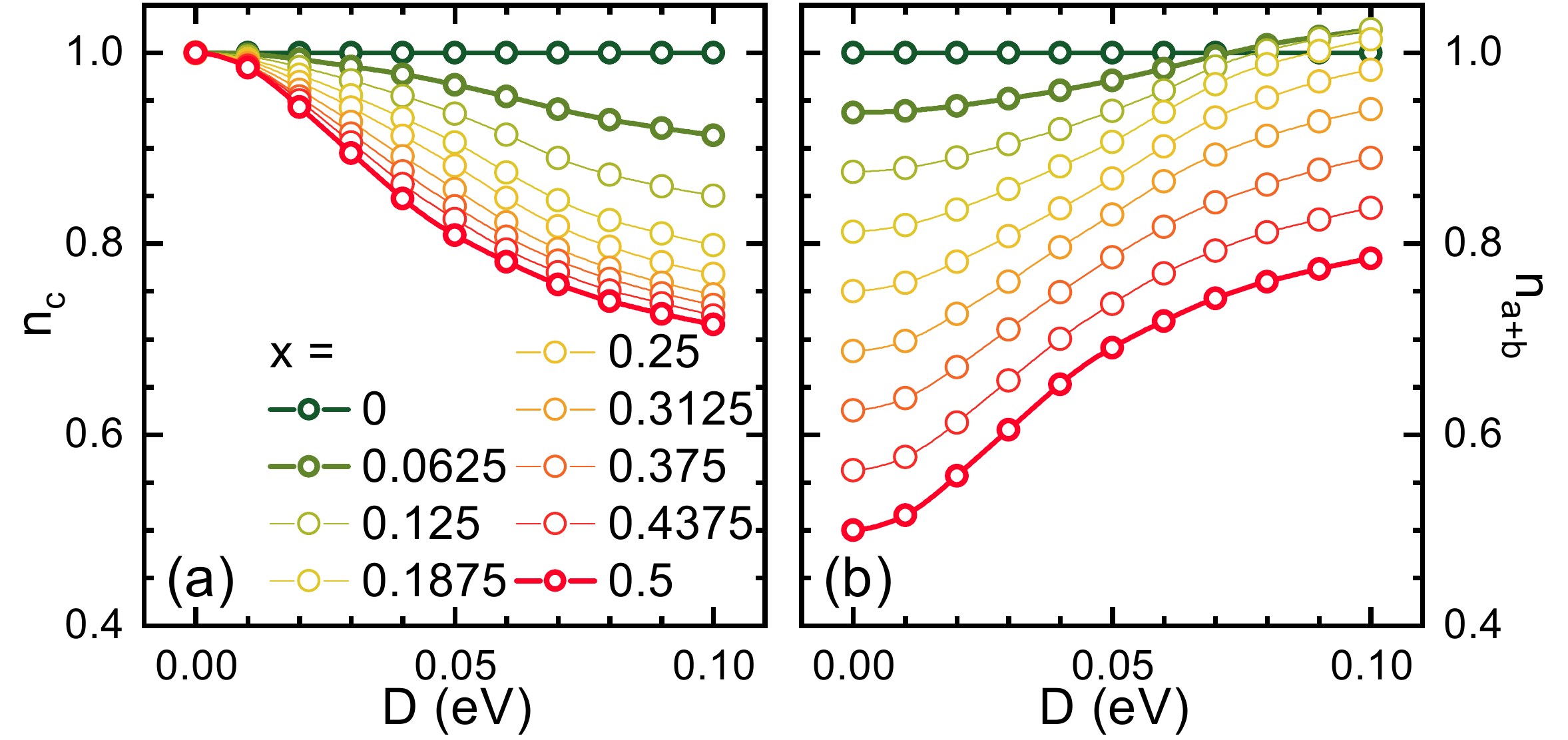}
\protect\caption{
Average electron density (per V ion) versus
orbital-polarization parameter ${\cal D}$ (\ref{pol}) for
doping $x\in[0.0,0.5]$ [legend in~(a)] for:
(a) $c$ orbitals, $n_c$;
(b) $\{a,b\}$ orbital doublet, $n_{a+b}$.
Parameters: $U=4.5$, $J_H=0.5$, $t=0.2$, $V_{\rm D}=2.0$ (all in eV).}
\label{fig:n}
\end{figure}

Each Ca$^{2+}$ defect injects a hole that replaces the $b'$ electron
on a defect cube with the highest energy in the actual defect
realization \cite{Ave18},
see Fig. \ref{fig:orbs}(b). Which V ion this is depends on the
interaction $v(r)$ with all other random defects and doped holes.
The unrestricted Hartree-Fock (UHF) method is well designed to study
spin-orbital order \cite{Miz95,Miz99,Miz01,Noh05}. The subtle
self-consistency problem, with random charged defects, is solved here
using the rotationally invariant UHF method, which is able to reproduce
the gap between the lower Hubbard band (LHB) and the UHB (with its
multiplet structure) for the perovskite vanadates \cite{Ave18}.
Statistical averages are performed over $M=100$ defect realizations,
and we have verified that, for the quantities presented here, it
suffices to consider \mbox{$N=4\times 4\times 4$}-size clusters.

In Fig.~\ref{fig:n}, we show how orbital polarization $\mathcal{D}$
influences charge densities $n_c$ and $n_{a+b}\equiv n_a+n_b$ for
increasing
doping $x$, where $n_{\alpha}=\langle\hat{n}_{\alpha}\rangle$ and
$\hat{n}_{\alpha}=\frac{1}{N}\sum_{i\sigma}\hat{n}_{i\alpha\sigma}$.
The case $\mathcal{D}=0$ is straightforward: doped holes go into the
higher lying $ab$ states, i.e., $n_{a+b}\!=\!1-x$ and $n_c\!=\!1$.
At finite $\mathcal{D}$, electrons occupy the rotated $|c'\rangle$
and $|b'\rangle$ orbitals that, for increasing $\mathcal{D}$, leads to
a decrease of $n_c$ and to an increase of $n_{a+b}$, which may even
exceed 1. This redistribution is evident in the large $\mathcal{D}$
limit where the occupied states become
\mbox{$|c'\rangle=(2|c\rangle-|a\rangle-|b\rangle)/\sqrt{6}$} and
\mbox{$|b'\rangle=(|a\rangle-|b\rangle)/\sqrt{2}$} for a V ion in (111)
position, see Fig. \ref{fig:orbs}(b), leading --- for small $x$ and
$t=0$ --- to occupations $n_{a+b}\!=\!1-x+\frac{8}{3}x$ and
$n_c\!=\!1-\frac{8}{3}x$.

\begin{figure}[t!]
\includegraphics[width=\columnwidth]{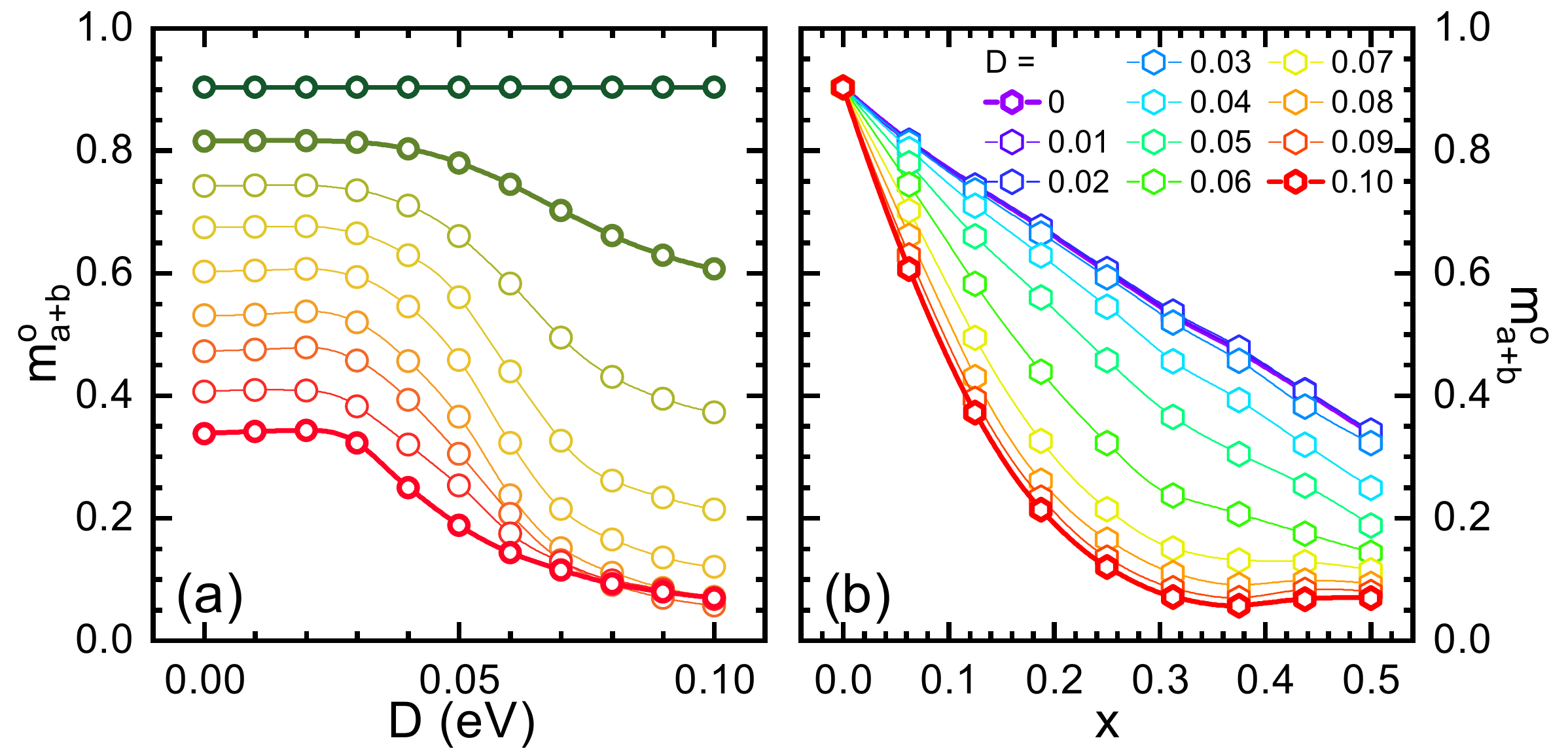}
\protect\caption{Orbital order parameter $m_{a+b}^o$ (\ref{oop}):
(a)~for increasing orbital polarization ${\cal D}$ at
different doping [same legend as in Fig. \ref{fig:n}(a)] and
(b)~for~increasing doping $x\in[0,0.5]$ at representative values
of ${\cal D}$ (see legend). Parameters as in Fig.~\ref{fig:n}.}
\label{fig:oop}
\end{figure}

The rotation of $t_{2g}$ orbitals reduces the OO parameter
describing the staggered $a/b$ order on each defect cube:
\begin{equation}
m_{a+b}^o\equiv\frac{1}{M}\sum_{s=1}^M\frac{1}{N}\sum_i\left\langle
\hat{n}_{ia}-\hat{n}_{ib}\right\rangle_s e^{i\mathbf{Q}_G\cdot\mathbf{R}_i},
\label{oop}
\end{equation}
where $\mathbf{Q}_G=(\pi,\pi,\pi)$ is the vector corresponding to the
\mbox{$G$-AO} order. One finds $m_{a+b}^o\simeq 0.9$ in the undoped 
case, see Fig. \ref{fig:oop}(a), i.e., due to the finite hopping 
$t=0.2$ eV. For $\mathcal{D}=0$, the order parameter
$m_{a+b}^o$ decreases almost linearly with $x$. This case has been
studied in a polaron theory using a small $t$ expansion \cite{Ave18}
where \mbox{$m_{a+b}^o\simeq1-x(1+2\delta_c)$}. The $1-x$ describes 
the dilution of electrons in $a$ or $b$ orbitals upon doping. The 
polarity parameter $\delta_c$ is 0 if the doped hole is localized on 
a single V site, and is finite, but less than 0.5, if it moves in a 
double exchange process along an \textit{active bond} (AB) \cite{Ave13}, 
thereby generating orbital defects. It is clear that the kinetic energy 
of holes in the $\mathcal{D}=0$ case \cite{Ave18} weakens the OO, but 
does not collapse it. In contrast, the $\mathcal{D}$ dependence in Fig. 
\ref{fig:oop}(a) is, for small $\mathcal{D}\le 0.03$ eV, almost absent 
and followed by a decay centered at $\mathcal{D}_c\approx 0.05$. 
We identify the orbital polarization interaction $\propto\mathcal{D}$ 
as the driving force of the decay.
For large doping $x\ge x_c\approx 0.3$ and $\mathcal{D}\ge 0.07$ eV,
there is a saturation of $m_{a+b}^o$ induced by the large number of
overlapping defect cubes. Simultaneously $C$-AF spin order persists in 
the regime where the OO melted. This behavior agrees with experimental 
data \cite{Tok00,Fuj05,Fuj08}.

We remark that the Hamiltonian parameters used here are relevant for
La$_{1-x}$Sr$_x$VO$_3$, where $G$-AO order disappears at 
$x_{\rm exp}\simeq 0.18$ \cite{Tok00}. It is worth noticing that the 
decay of the OO is due to a field term in the Hamiltonian, which 
explains its rather gradual decline, a trend also seen in experiments 
\cite{Fuj05,Ree16}. So far, we have not observed in our data 
the collective features expected for conventional phase transitions.

\begin{figure}[t!]
\includegraphics[width=.96\columnwidth]{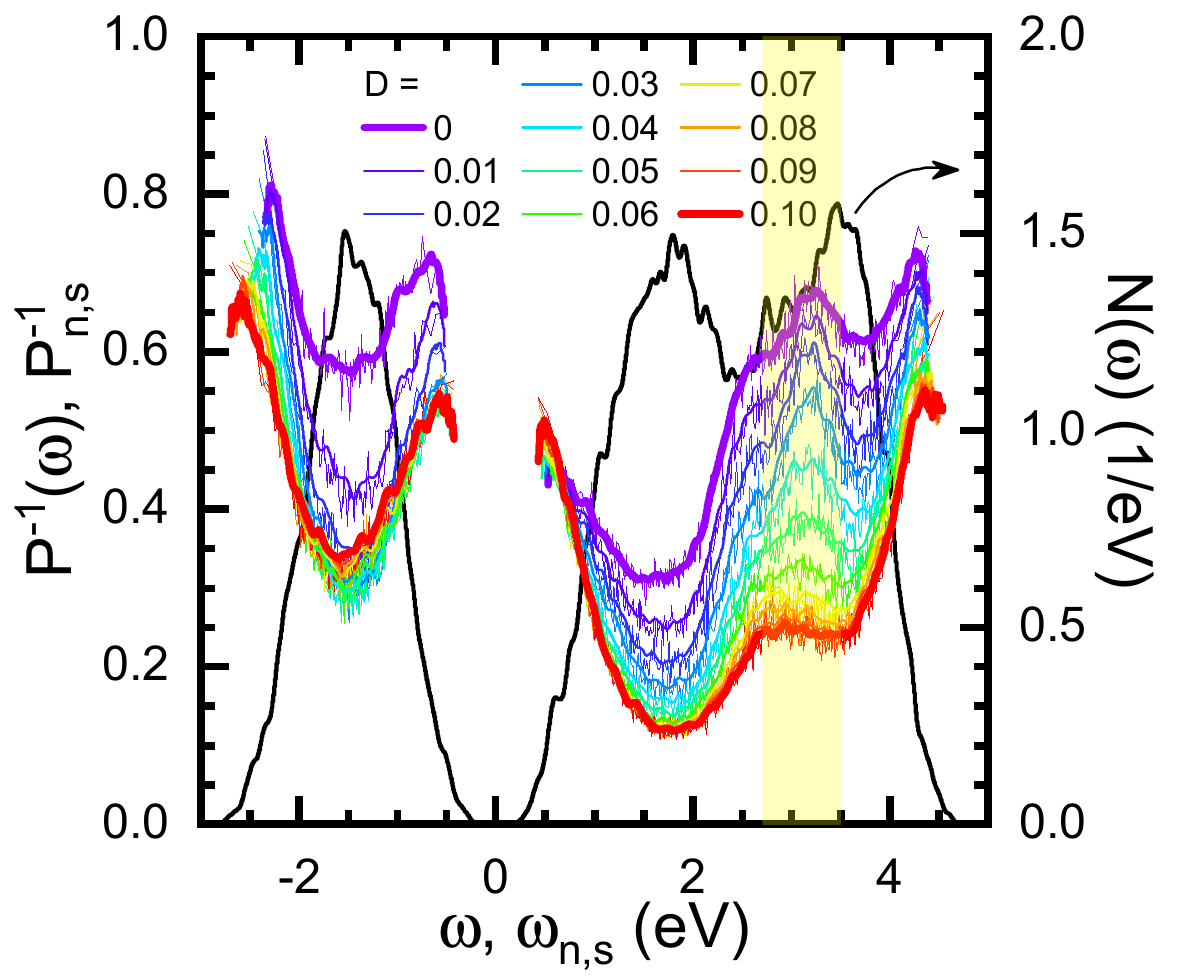}
\protect\caption{
IPN spectrum $P^{-1}_{n,s}$ versus $\omega_{n,s}$ and average
$P^{-1}(\omega)$ for different $\mathcal{D}$ (see legend) at
\mbox{$x=0.3125$}. Black lines show $N\left(\omega\right)$ with
LHB/UHB for $\mathcal{D}=0$. The $\mathcal{D}$ dependence of HS
$d^2\!\rightarrow\!d^3$ transitions on defect cubes at
\mbox{$\omega\in(2.7,3.5)$} eV, reflects the melting of OO
(see shading). Parameters as in Fig.~\ref{fig:n}.
}
\label{fig:iPN}
\end{figure}

Given the randomness of these systems, how does the localization of
states change with orbital polarization \mbox{$\propto\mathcal{D}$?}
A convenient measure of the degree of localization of a UHF wave
function $\psi_{n,s}(r)$ is the participation number (PN) $P_{n,s}$,
which is 1 for a state localized on a single site and $N$ for a Bloch
state. Usually, one considers the inverse participation number (IPN)
which takes the form \cite{Ave18}, \mbox{$P_{n,s}^{-1}=\sum_i(
\sum_{\alpha\sigma}|\langle\psi_{n,s}|i\alpha\sigma\rangle|^2)^2\in[0,1]$}
for systems with spin-orbital degeneracy. $P_{n,s}^{-1}$ is plotted in Fig.
\ref{fig:iPN} for \mbox{$x=0.3125$} versus the respective eigenvalues
$\omega=\omega_{n,s}$ for all $6N$ states $n$ and \mbox{$M=100$} defect
realizations $s$ together with the average IPN spectra $P(\omega)^{-1}$
\cite{Ave18}. Interestingly, despite the strong
changes in the UHF wave functions $\psi_{n,s}(r)$, the density of states
\mbox{$N\left(\omega\right)\equiv\frac{1}{M}\sum_{s=1}^{M}\left[
\frac{1}{N}\sum_{n=1}^{6N}\delta(\omega-\omega_{n,s})\right]$} hardly
changes for $\mathcal{D}\le 0.1$ eV; thus we show it only for
$\mathcal{D}=0$. Overall, one
recognizes a gradual decrease of the IPN values with increasing
$\mathcal{D}$ and a saturation for $\mathcal{D}\ge 0.08$ eV, where the
OO is practically absent. The PN results in maximum $3$ ($8$) sites
for the LHB (UHB): all states remain well localized. The discontinuity
of the IPN at $\mathcal{D}=0$ between removal and addition states,
right below and above the Fermi energy $\mu$, has been discussed
before \cite{Ave18}. Here, we observe its disappearance at moderate
$\mathcal{D}$: delocalization of removal states can be attributed to
the orbital rotation leading to $c'$ orbital \cite{note1}.

For the $\mathcal{D}$ dependence of the IPN (Fig.~\ref{fig:iPN}), the
energy interval \mbox{$\omega\in(2.7,3.5)$} eV is special and shows
the largest variation in the range \mbox{$0.04<\mathcal{D}<0.07$} eV,
similar to the $\mathcal{D}$ dependence of $m_{a+b}^o$ for $x=0.3125$,
shown in Fig. \ref{fig:oop}(a).
There are three different types of $d^2\rightarrow d^3$ transitions
that fall into this energy window. Namely, either one of the two low
spin (LS) transitions in the host or the high spin (HS) transition on
a defect cube, where excitation energies are increased by $V_D$, i.e.,
\mbox{$\omega_{\rm HS}\!=U-3J_H+V_D+\omega_{\rm LHB}\approx 3.0$} eV,
where the position of the LHB is given by 
\mbox{$\omega_{\rm LHB}\!=E_{\rm LHB}-\mu\approx-V_D$} \cite{Ave18}.
It is the $\omega_{\rm HS}$ transitions that are sensitive to the 
melting of OO.

\begin{figure}[t!]
\includegraphics[width=\columnwidth]{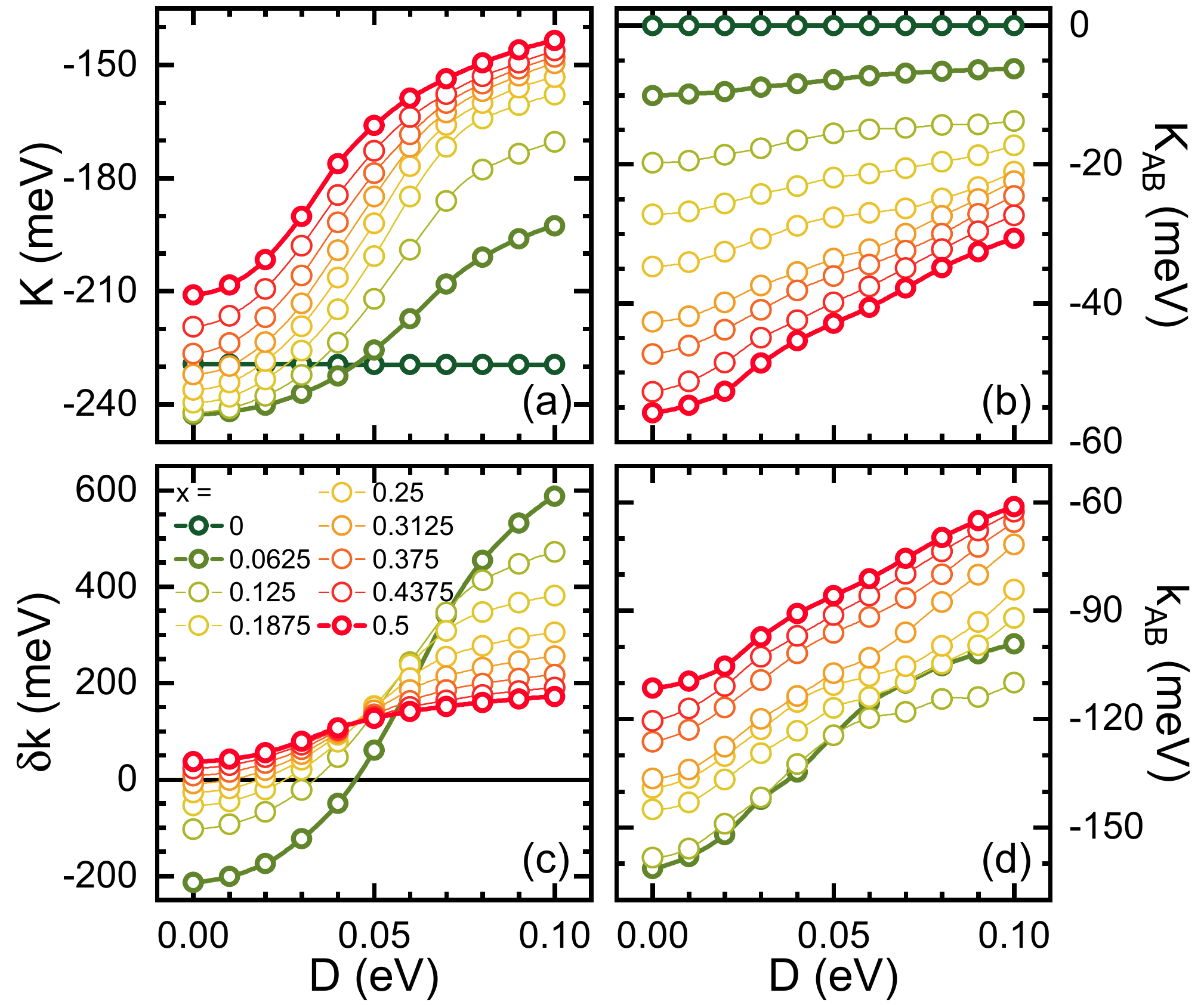}
\protect\caption{
Top --- Kinetic energies per site versus orbital polarization strength
${\cal D}$ for different doping $x\in[0,0.5]$
[for color convention see legend in (c)]:
(a)~total energy $K$, and
(b) the kinetic energy on active bonds $K_{\rm AB}$.
Bottom --- Energies per defect (\ref{deK}):
(c) change of total kinetic energy $\delta k$  (relative to $x=0$), and
(d) the kinetic energy on an active bond $k_{\rm AB}$.
Parameters as in Fig. \ref{fig:n}.}
\label{fig:kin}
\end{figure}

We have discussed above that the rotation of orbitals lowers the
Coulomb energy of electrons in the electric field of defects. So far,
we have not explained which mechanism opposes the rotation and
determines the characteristic scales $\mathcal{D}_c$ and $x_c$ in Fig.
\ref{fig:oop}. We show here that both are indeed determined by the
kinetic energy of the system rather than by the CF
--- a consequence of strong correlations. First, we analyze in Fig.
\ref{fig:kin}(a) the total kinetic energy per vanadium site,
\mbox{$K(x)\equiv\frac{1}{N}\langle\tilde{\cal H}_{\rm kin}\rangle$},
which includes both the hopping $\propto t$ and Fock
$\propto v(r_{ij})$ terms \cite{Ave13}.
For the undoped Mott insulator ($x=0$), such as LaVO$_3$, we find large
kinetic energy $K(0)\simeq -230$ meV, see the horizontal $x=0$ line in
Fig. \ref{fig:kin}(a). This is equivalent to the sum of the 
spin-orbital superexchange energies for the three cubic bond directions
\cite{Ole05}. For all other $x\ge 0.0625$, one
finds a monotonic increase of $K$ (i.e., loss of superexchange) for
increasing either $x$ or $\mathcal{D}$. Note the complementary trends
in the decay of the OO parameter $m_{a+b}^o$ in Fig. \ref{fig:oop}(a).

From a polaron perspective, the \textit{increase} of $K$ is puzzling as
one may expect that added holes would lead to delocalization, giving
rise to some extra negative kinetic energy. In fact, for small
$\mathcal{D}$ and $x$, the kinetic energy $K$ in Fig. \ref{fig:kin}(a)
is indeed lower than the energy of the undoped system $K(0)$, in
agreement with intuition. The dominant kinetic energy gain is expected
to stem from $d^2d^1\rightarrow d^1d^2$ double exchange process on  
active FM bonds as confirmed by looking at the total UHF kinetic
energy of holes on ABs $K_{\rm AB}$, see Fig. \ref{fig:kin}(b).
We also consider the kinetic energy gain per defect $\delta k$ and per
active bond $k_{\rm AB}$, or equivalently per doped hole,
\begin{equation}
\delta k\equiv \left[K(x)-K(0)\right]/x, \hskip .3cm
k_{\rm AB}\equiv K_{\rm AB}(x)/x,
\label{deK}
\end{equation}
The kinetic energy gain $\delta k$ shown in Fig. \ref{fig:kin}(c)
reveals an approximate isosbestic point, where $\delta k$ increases
(decreases) as function of $x$ for small (large) $\mathcal{D}$.
For $\mathcal{D}=0$ in the dilute case \mbox{($x=0.0625$)}, the
kinetic energy gain is \mbox{$\delta k=-0.208$} eV, while the kinetic
energy of a hole on an active bond in Fig. \ref{fig:kin}(d) is
$k_{\rm AB}\approx -0.162 $ eV. To better appreciate these numbers we
recall that $t=0.2$ eV. Thus, we conclude that $k_{\rm AB}$ is in fact
the dominant contribution of the total kinetic energy gain $\delta k$
at $x=0.0625$ and small $\mathcal{D}$. For larger doping and small
$\mathcal{D}$, the kinetic energy per hole is quenched due to
electron-electron and electron-defect interactions \cite{Ave18},
and the formation of localized bipolarons (ABs with 2 doped holes)
created by touching defect cubes \cite{suppl}.

Next, we turn to the $\mathcal{D}$ dependence of $\delta k$ and
$k_{\rm AB}$ in Figs. \ref{fig:kin}(c) and \ref{fig:kin}(d). For low
(high) doping $x=0.0625$ (0.50), the change of $\delta{k}$ between
$\mathcal{D}=0$ and 0.1 eV is 800 (200) meV, i.e., much more than
the change of $k_{\rm AB}$ which is only 60 (40) meV. This clearly
shows that the $\mathcal{D}$-dependent change of $\delta k$ is mainly
due to the orbital rotation at all corners of the defect cube and not
just at the active bond. The smaller values at high doping result from
the frustration of orbital rotation due to the touching of defect
cubes. In view of the significant overlap of defect cubes at already
moderate doping, one may expect that some states extend over several 
cubes. Yet, in the analysis of IPN we have shown in Fig. 
\ref{fig:iPN} that such delocalized states do not exist and holes 
injected into the LHB do extend typically just over two to three V 
sites.

Summarizing, we have shown that the dominant mechanism that leads to
the collapse of the orbital order is not the motion of doped holes, 
but the orbital rotation induced by charged defects on their vanadium 
neighbors. This field induced suppression of the orbital order is 
non-cooperative and does not lead to a conventional phase transition, 
--- like the loss of antiferromagnetic order in high-$T_c$ cuprates 
\cite{Kha93}. We believe that our model gives a qualitative explanation 
of the decay of the orbital order accompanied by robustness of spin 
order in $R_{1-x}$Sr$_x$VO$_3$ compounds.

%%%%%%%%%%%%%%%%%%%%%%%%%%%%%%%%%%%%%%%%%%%%%%%%%%%%%%%%%%%%%%%%%%%%%%%%%%%%%%
\acknowledgments
%%%%%%%%%%%%%%%%%%%%%%%%%%%%%%%%%%%%%%%%%%%%%%%%%%%%%%%%%%%%%%%%%%%%%%%%%%%%%%
%\textit{Acknowledgements:}
A. M. O. acknowledges Narodowe Centrum Nauki
(NCN, Poland) Project No. 2016/23/B/ST3/00839 and is 
\mbox{grateful} for the Alexander von Humboldt Foundation 
Fellowship (Humboldt-Forschungspreis).

\appendix*
\section{Supplemental Material}

In the first Section of this Supplemental Material, we present the
three-band extended Hubbard model used in the main text, which 
describes the $t_{2g}$ electronic states in vanadium perovskites. 
In Section II, we explain in detail how the general form of the 
orbital polarization interaction, describing the rotation of $t_{2g}$ 
orbitals on the vanadium neighbors of charged defects, is obtained 
from the representative term acting on the $\{a,b\}$ orbital doublet 
reported in the main text. In Section III, we show that spin order 
decouples from orbital order and is robust at increasing doping; 
it is not influenced by orbital polarization interaction.
Finally, in Section IV, we discuss some of the interaction and 
frustration effects appearing on increasing defect density where 
more and more \emph{defect cubes} have common corners, edges or faces.

%%%%%%%%%%%%%%%%%%%%%%%%%%%%%%%%%%%%%%%%%%%%%%%%%%%%%%%%%%%%%%%%%%%%%%%

\subsection{The three-band Hubbard model}

%%%%%%%%%%%%%%%%%%%%%%%%%%%%%%%%%%%%%%%%%%%%%%%%%%%%%%%%%%%%%%%%%%%%%%%

The Hamiltonian for $t_{2g}$ electrons in doped vanadium 
(La,Y)$_{1-x}$Ca$_{x}$VO$_{3}$ perovskites \cite{Hor11,Ave13}, 
\begin{equation}
{\cal H}_{t2g}\!={\cal H}_{{\rm Hub}}+{\cal H}_{{\rm pol}}
+\sum_{i<j}v(r_{ij})\hat{n}_{i}\hat{n}_{j}
+\sum_{mi}v(r_{mi})\hat{n}_{i},\label{Ht2gs}
\end{equation}
includes the three-band Hubbard model ${\cal H}_{{\rm Hub}}$ 
\cite{Ole07} for the reference host system without charged defects. It 
acts on the
electrons in $t_{2g}$ orbital states similar to the model for pnictides
\cite{Dag11} and consists of the kinetic energy ${\cal H}_{{\rm kin}}$,
local interactions described by the degenerate Hubbard model 
${\cal H}_{{\rm U-J_{\mathrm{H}}}}$, supplemented by rather weak terms: 
the CF splitting ${\cal H}_{{\rm CF}}$,
and the JT interactions ${\cal H}_{{\rm JT}}$,
\begin{equation}
{\cal H}_{{\rm Hub}}={\cal H}_{{\rm kin}}
+{\cal H}_{{\rm U-J_{\mathrm{H}}}}+{\cal H}_{{\rm CF}}
+{\cal H}_{{\rm JT}}.\label{3band}
\end{equation}

The kinetic energy reads as, 
\begin{equation}
{\cal H}_{{\rm kin}}=
\sum_{{\langle ij\rangle\parallel\gamma\atop \alpha\sigma}}
t_{ij}^{\gamma\alpha}\left(\hat{d}_{i\alpha\sigma}^{\dagger}\hat{d}_{j\alpha\sigma}^{}
+\hat{d}_{j\alpha\sigma}^{\dagger}\hat{d}_{i\alpha\sigma}^{}\right).
\label{Hts}
\end{equation}
Here, $\hat{d}_{i\alpha\sigma}^{\dagger}$ is the electron creation operator
in the $t_{2g}$ orbitals $\alpha\in\{xy,yz,zx\}$ with spin 
$\sigma=\uparrow,\downarrow$
at site $i$. The effective hopping $t_{ij}^{\gamma\alpha}$ of $t_{2g}$
electrons between two vanadium ions at sites $i$ and $j$ depends
on bond direction $\langle ij\rangle\parallel\gamma$ and on the orbital
flavor $\alpha$. It occurs via hybridization with an intermediate
oxygen $2p_{\pi}$ orbital along 180° V--O--V bonds. Therefore,
the hopping: (i) is diagonal and conserves the orbital flavor $\alpha$
when $\alpha\!\neq\!\gamma$ and hybridization is finite, i.e., 
$t_{ij}^{\gamma\alpha}=-t$,
and (ii)~vanishes in one of the three cubic directions for which
the hybridization with oxygen $2p_{\pi}$ orbitals vanishes by symmetry,
i.e., $t_{ij}^{\gamma\gamma}=0$. Using these properties, it is convenient
to introduce the following short-hand notation for the orbital degree
of freedom \cite{Kha00}, 
\begin{equation}
|a\rangle\equiv|yz\rangle,\hskip.7cm
|b\rangle\equiv|zx\rangle,\hskip.7cm
|c\rangle\equiv|xy\rangle,\nonumber
%\label{t2g}
\end{equation}
with the labels $\gamma=a,b,c$ referring to the cubic axis along
which the hopping element vanishes.

Local interactions at vanadium ions are described by the degenerate
Hubbard model ${\cal H}_{{\rm U-J_{\mathrm{H}}}}$ parametrized by
two Kanamori parameters: intraorbital Coulomb interaction $U$ and
Hund's exchange $J_{H}$ between two $t_{2g}$ electrons \citep{Ole83},
\begin{align}
{\cal H}_{{\rm U-J_{\mathrm{H}}}} & =U\sum_{i\alpha}\hat{n}_{i\alpha\uparrow}
\hat{n}_{i\alpha\downarrow}+J_{H}\!\sum_{i,\alpha\neq\beta}
\hat{d}_{i\alpha\uparrow}^{\dagger}\hat{d}_{i\alpha\downarrow}^{\dagger}
\hat{d}_{i\beta\downarrow}^{}\hat{d}_{i\beta\uparrow}^{}\nonumber \\
& +\sum_{i,\alpha<\beta}\!\left[\left(U-\frac{5}{2}J_{H}\right)
\hat{n}_{i\alpha}\hat{n}_{i\beta}
-2J_{H}\hat{\vec{S}}_{i\alpha}\!\cdot\!\hat{\vec{S}}_{i\beta}\right]\!.
\end{align}
Interorbital Coulomb interactions $\propto n_{i\alpha}n_{i\beta}$
are expressed in terms of spin-orbital electron density operators,
\mbox{$\hat{n}_{i\alpha}^{}=\sum_{\sigma}\hat{n}_{i\alpha\sigma}^{}=
\sum_{\sigma}\hat{d}_{i\alpha\sigma}^{\dagger}\hat{d}_{i\alpha\sigma}^{}$};
orbital spin operators, 
\mbox{$\hat{\vec{S}}_{i\alpha}\equiv\{
\hat{S}_{i\alpha}^x,\hat{S}_{i\alpha}^y,\hat{S}_{i\alpha}^z\}$},
appear in the Hund's exchange  
\mbox{$\propto-J_{H}\hat{\vec{S}}_{i\alpha}\!\cdot\!\hat{\vec{S}}_{i\beta}$.}
In a Mott insulator, charge fluctuations are quenched and electrons
localize due to large $U\gg t$. In case of LaVO$_{3}$, one finds
a $t_{2g}^{2}$ configuration at each vanadium ion and Hund's exchange
$J_{H}$ stabilizes high spin states with $S=1$. The insulating ground
state of LaVO$_{3}$ has a $C$-type antiferromagnetic ($C$-AF) spin
coexisting with $G$-type alternating orbital ($G$-AO) order 
\citep{Ole07}. 

The structural transition at $T_{s}\sim200$ K lifts the degeneracy
of the three $t_{2g}$ orbitals and breaks the cubic symmetry in the
orbital space \cite{Ole07}. At low temperature, the CF splitting favors 
$xy\equiv c$ orbitals by energy $\Delta_{c}=0.1$ eV, which we take as a 
constant parameter independent of temperature, and the CF Hamiltonian 
is, 
\begin{equation}
{\cal H}_{{\rm CF}}=-\Delta_{c}\sum_{i}\hat{n}_{ic}.
\label{Hcfs}
\end{equation}
It selects the orbital doublet as orbital degree of freedom and gives 
either $c_{i}^{1}a_{i}^{1}$ or $c_{i}^{1}b_{i}^{1}$ configuration at 
the V ion sitting at site $i$, depending on the actual lattice 
distortion in the $ab$ plane. In a Mott insulator, spin-orbital 
superexchange explains the ground state observed in LaVO$_{3}$ 
\cite{Kha01}.

Lattice distortions change the electronic state and
induce weak JT interactions in the three-band model (\ref{3band}),
\begin{align}
{\cal H}_{\rm JT}&=\frac{1}{4}\,V_{ab}\sum_{\langle ij\rangle{\parallel}ab}
(\hat{n}_{ia}-\hat{n}_{ib})(\hat{n}_{ja}-\hat{n}_{jb})\nonumber \\
& -\frac{1}{4}\,V_{c}\sum_{\langle ij\rangle{\parallel}c}
(\hat{n}_{ia}-\hat{n}_{ib})(\hat{n}_{ja}-\hat{n}_{jb}).
\end{align}
Using the orbital $\tau_{i}^{z}$ operators, 
\begin{equation}
\tau_{i}^{z}\equiv
\frac{1}{2}\sum_{\sigma}\left(\hat{d}_{ia\sigma}^{\dagger}\hat{d}_{ia\sigma}^{}
-\hat{d}_{ib\sigma}^{\dagger}\hat{d}_{ib\sigma}^{}\right),\label{tauzs}
\end{equation}
the JT interactions are, 
\begin{equation}
{\cal H}_{{\rm JT}}=
V_{ab}\sum_{\langle ij\rangle{\parallel}ab}\hat{\tau}_i^z\hat{\tau}_j^z
-V_{c}\sum_{\langle ij\rangle{\parallel}c }\hat{\tau}_i^z\hat{\tau}_j^z.
\end{equation}
These interactions stabilize another competing type of spin-orbital
order \cite{Kha01}, the $G$-type AF ($G$-AF) spin coexisting with 
$C$-type AO ($C$-AO) order, which represents the ground state in YVO$_3$ 
\cite{Hor11,Ren00}. Small doping $x\simeq 0.01$ leads to a phase 
transition to the $C$-AF/$G$-AO phase, which is the phase studied in 
this work.

Following the earlier studies, we have fixed the small parameters in 
${\cal H}_{\rm CF}$ and ${\cal H}_{\rm JT}$ as follows: $\Delta_c=0.1$,
$V_{ab}=0.03$, and $V_{c}=0.05$ (all in eV). The term $\propto V_{ab}$
favors alternating $\{a,b\}$ orbitals, i.e., AO order in the $ab$
planes ($V_{ab}>0$) while the \emph{ferro-orbital} order is favored
along the $c$ cubic axis ($V_{c}>0$). Thus, the term $\propto V_{c}$
weakens the superexchange orbital interaction $\propto Jr_{1}$, where
$J=4t^{2}/U$ and $r_{1}=(1-3\eta)^{-1}$ with $\eta=J_{H}/U$, which
along the $c$ axis favors the observed $G$-AO order \cite{Kha01}.
One finds that for the present parameters ($U=4.5$, $t=0.2$, $J_{H}=0.5$,
all in eV) $Jr_{1}=53$ meV, so taking $V_{c}=50$ meV one is indeed
close to the switching of the orbital order observed in YVO$_{3}$ 
\cite{Fuj10,Sah17}.

%%%%%%%%%%%%%%%%%%%%%%%%%%%%%%%%%%%%%%%%%%%%%%%%%%%%%%%%%%%%%%%%%%%%%%%

\subsection{Orbital polarization around defects}

%%%%%%%%%%%%%%%%%%%%%%%%%%%%%%%%%%%%%%%%%%%%%%%%%%%%%%%%%%%%%%%%%%%%%%%

%%%%%%%%%%%%%%%%%%%%%%%%%%%%%%%%%%%%%%%%%%%%%%%%%%%%%%%%%%%%%%%%%%%%
%%                           Table I
%%%%%%%%%%%%%%%%%%%%%%%%%%%%%%%%%%%%%%%%%%%%%%%%%%%%%%%%%%%%%%%%%%%%
\begin{table}[b!]
\caption{The coefficients $\lambda_{\alpha\beta}(\mathbf{r}_{i}-\mathbf{R}_{m})$
in Eq. (\ref{pol}) for different orbital doublets $\{\alpha,\beta\}$
and for different directions $\{(\mathbf{r}_{i}-\mathbf{R}_{m})\}$.}
\vskip .1cm 
\begin{ruledtabular}
\begin{tabular}{cccccc}
orbital doublet  & 
\multicolumn{4}{c}{$(\mathbf{r}_{i}-\mathbf{R}_{m})\!\parallel$} & 
\tabularnewline
$\{\alpha,\beta\}$ & $(111)$ & $(11\bar{1})$ & $(\bar{1}11)$ & $(1\bar{1}1)$ & \tabularnewline
\colrule 
$\{a,b\}$ &  $1$  & $\,\;\;1$ &   $-1$    &     $-1$  & \tabularnewline
$\{a,c\}$ &  $1$  &   $-1$    &   $-1$    & $\,\;\;1$ & \tabularnewline
$\{b,c\}$ &  $1$  &   $-1$    & $\,\;\;1$ &     $-1$  & \tabularnewline
\end{tabular}
\end{ruledtabular}

\label{tab:para} 
\end{table}

The orbital polarization term results from electron-defect interaction
and modifies the orbital basis at V ions on the \emph{defect cube}
$\mathcal{C}_{m}$ around the charged defect at $\mathbf{R}_{m}$
\cite{Ave13}: 
\begin{equation}
\mathcal{H}_{\rm pol}=\mathcal{D}\!\!
\sum_{m,i\in\mathcal{C}_m \atop \alpha\neq\beta,\sigma}\!\!
\lambda_{\alpha\beta}(\mathbf{r}_{i}-\mathbf{R}_{m})\!\left(
 \hat{d}_{i\alpha\sigma}^{\dag}\hat{d}_{i \beta\sigma}^{} 
+\hat{d}_{i \beta\sigma}^{\dag}\hat{d}_{i\alpha\sigma}^{}\right)\!.
\label{pols}
\end{equation}
The coefficients $\lambda_{\alpha\beta}(\mathbf{r}_i\!-\mathbf{R}_m)=\pm1$
are selected to minimize the Coulomb repulsion with the defect charge.
Taking the $\{a,b\}$ doublet active along the $c$ axis as an example,
one finds that $\lambda_{ab}\left(\mathbf{r}_i\!-\!\mathbf{R}_m\right)=+1$
for the directions 
$\left(\mathbf{r}_i\!-\!\mathbf{R}_m\right)\!\parallel\!(111)$,
$\left(\mathbf{r}_i\!-\!\mathbf{R}_m\right)\!\parallel\!(11\bar{1})$,
and $-1$ for the other two diagonal directions \cite{Hor11}, see 
Table I. For this doublet the eigenstates of the polarization operator,
\begin{equation}
\tau_{i}^{x}\equiv\frac{1}{2}\sum_{\sigma}\left(
 \hat{d}_{ia\sigma}^{\dagger}\hat{d}_{ib\sigma}^{}
+\hat{d}_{ib\sigma}^{\dagger}\hat{d}_{ia\sigma}^{}\right),
\label{taux}
\end{equation}
at site $i$ have energy either lowered or increased by $\mathcal{D}$,
depending on whether they are directed towards the defect site $m$
or have lobes in the plane being orthogonal to the above direction, 
see Fig. 1 (main text). 

The remaining values of 
$\lambda_{\alpha\beta}\left(\mathbf{r}_{i}-\mathbf{R}_{m}\right)$,
which determine the local mixing of $\{a,c\}$ or $\{b,c\}$ orbitals
in analogy to Eq. (\ref{taux}), may be obtained by simultaneous cyclic
permutations of the orbitals $\{a,b,c\}$ and of the cubic axes in the
direction of the vector $(\mathbf{r}_{i}-\mathbf{R}_{m})$, see Table~I.
Note that each direction along one of the cube's diagonals involves
two vanadium ions.

\subsection{Orbital polarization dependence of spin order}

The defect-induced orbital polarization (\ref{pol}) strongly affects 
the orbital order as well as the doping dependence of the $G$-type 
orbital order parameter $m^o_{ab}$, as we have shown in Fig. 3(b). 
Here, we present a complementary picture that shows how 
the decrease of spin order parameter, $m^s$, in the $C$-AF phase as a
function of doping $x$, changes with the orbital polarization strength 
$\mathcal{D}$. The spin order parameter is defined as,
\begin{equation}
m^s=\sum_{i,\nu} \left\langle 
 \hat{d}_{i\nu\uparrow}^{\dagger}\hat{d}_{i\nu\uparrow}^{}
-\hat{d}_{i\nu\downarrow}^{\dagger}\hat{d}_{i\nu\downarrow}^{}\right\rangle 
e^{i \mathbf{Q}_C\cdot\mathbf{r}_i},  
\end{equation}
where $\mathbf{Q}_C\equiv (\pi,\pi,0)$, and the sum is over all sites 
${\bf r}_i$ and orbital flavors \mbox{$\nu=a,b,c$}. 

\begin{figure}[b!]
\includegraphics[width=.9\columnwidth]{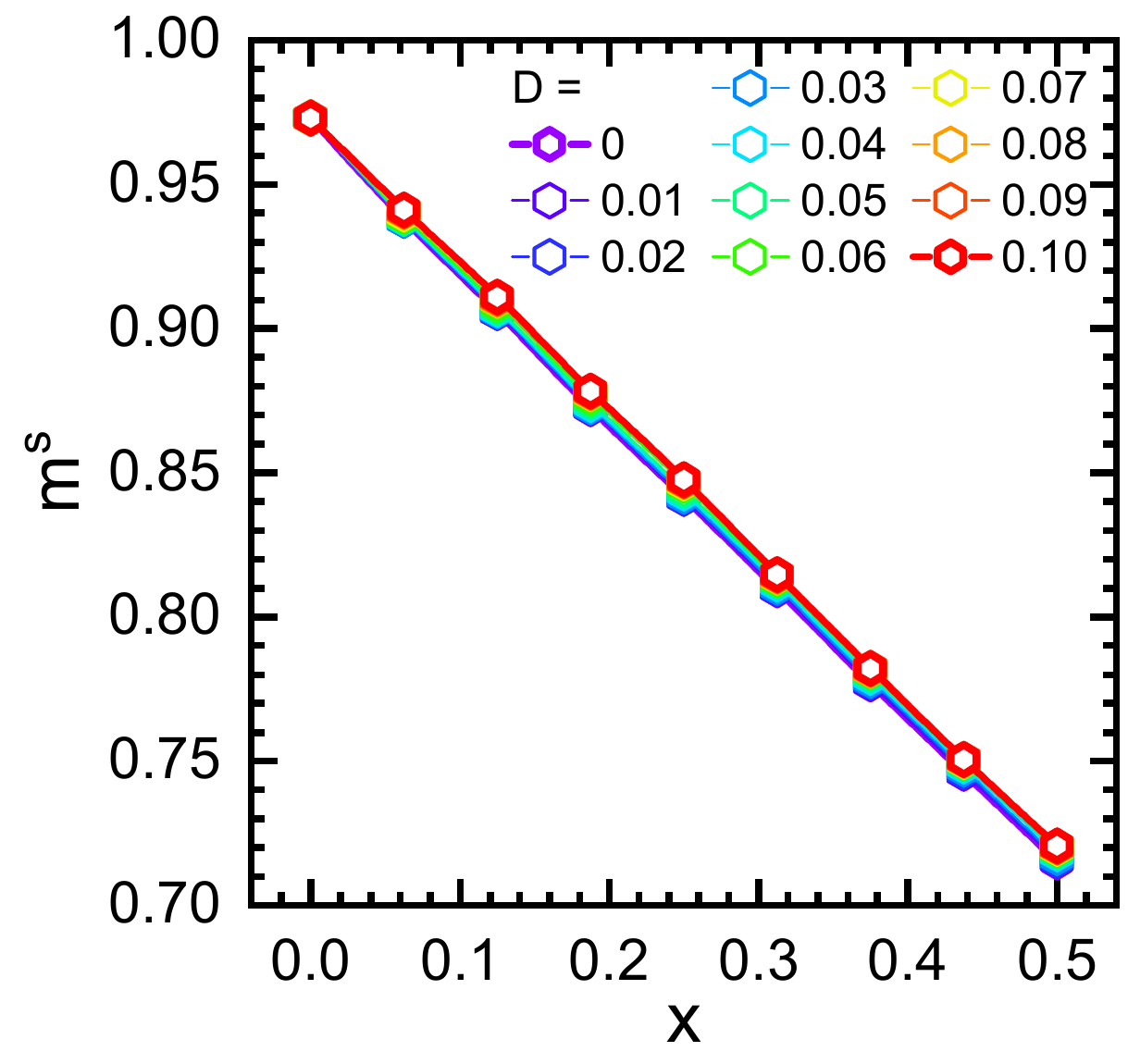}
\protect\caption{Spin-order parameter $m^s$ versus doping $x$ and its 
dependence on the orbital polarization strength $\mathcal{D}$.
Parameters as in the main text.}
\label{fig:oos}
\end{figure}

The results in Fig. \ref{fig:oos} show that the spin order parameter 
$m^s$ has an approximate linear decay with $x$, and only an extremely 
weak dependence on $\mathcal{D}$. The latter may be easily understood 
by the argument that orbital rotation does not affect spins. Yet, 
a so weak dependence on $\mathcal{D}$ is surprising if we go back to 
the origins of the decay with $x$. The latter has been explained in 
terms of a spin-polaron approach in Ref. \cite{Ave18}. There are two 
basic contributions to the almost linear decrease of $m^s$ with~$x$: 
\hfill\break
(a) the dilution of spins due to the added holes and \hfill\break
(b) the kinetic energy or string-formation effect due to the motion
of doped holes in the $C$-AF background. \hfill\break
Orbital rotation leads to the mixing of flavors and thus to the 
appearance of off-diagonal hopping processes that affect the kinetic 
energy. From this perspective, the insensitivity of the spin-order to 
orbital rotation, which we observe in Fig. \ref{fig:oop}, comes as a 
surprise. The solution of the puzzle follows from the observation
that holes form {\it small} spin-orbital polarons that are bound to 
defects where the kinetic energy string contributions are small.

We note that similar trends were reported in a recent 
experimental study of spin and orbital disordering by hole doping in 
Pr$_{1-x}$Ca$_x$VO$_3$ \cite{Ree16}. In that system, the long-range  
$C$-type ($C$-AF) spin order persists beyond the insulator-metal
transition crossover regime ($0.22<x<0.25$) and a N\'eel transition 
is still observed in the regime where the orbital order melted.

\subsection{Frustration resulting from overlapping defect cubes}

As we have argued in the text, the self-consistent UHF algorithm is
capable to obtain the electronic structure of the doped Mott insulator
even at high doping, where defect cubes share faces, edges or
just corners. The final results presented in the paper are averages 
over many defect realizations --- nevertheless one certainly would like 
to get some deeper insight or idea of the energy changes resulting from 
orbital rotations beyond the dilute limit, that is when frustration due 
to overlapping defect cubes is essential. This is possible by a careful 
analysis of correlation functions of individual random systems. In the 
following, we give for the interested reader a qualitative description 
of the most important effects due to the touching of defect cubes.

The increase with doping $x$ of the number of defect cubes 
sharing corners, edges, and faces  
is responsible for the non-linearity with doping $x$ of the behavior
of $n_{c}$, $n_{a+b}$ and $m_{a+b}^{o}$ at finite $\mathcal{D}$.
Actually, the percolation limit for defect cubes taking into
account simultaneously corner, edge and face sharing between V cubes
is counterintuitively low: $x_{p}=0.0976$ \cite{Domb}. This explains 
why $x=0.0625$ is somehow different from all other dopings we report. 
This also says that having defect cubes that share, 
in particular, one vertical bond is not that unusual also for very low 
values of doping. This brings in the possibility to confine two holes 
on that bond and gain substantial Coulomb potential energy (minimizing 
simultaneously the distances between the two holes and the two defects) 
at the expense of the kinetic energy gain we usually have at 
\emph{active bonds}. In other words, small spin-orbital polarons merge 
to give \emph{bipolarons}. This becomes more and more relevant for 
increasing values of $\mathcal{D}$ as active bonds gain less and less 
kinetic energy because of the induced orbital rotation. 

It is worth noting that two defect cubes sharing a face along 
$a$ or $b$ direction can simultaneously gain Coulomb potential energy 
and kinetic energy by confining the two holes over the two shared 
vertical bonds, but this leads again to two active bonds 
as if they would be on separate defect cubes \cite{Domb}. 
Also sharing just corners, horizontal edges and faces along the $c$ 
direction does not change the actual number of active bonds 
and has effects only on the potential energy one can gain.
Accordingly, only defect cubes sharing vertical bonds really
affects the kinetic energy of the system as they can reduce the 
overall number of active bonds.

Actually, the main source of kinetic energy loss with increasing $x$
is just such sharing of empty vertical bonds (occupied by a bipolaron). 
The loss of kinetic
energy by orbital rotation on all vertical bonds of a defect cube 
on increasing $\mathcal{D}$ is the other relevant source of
kinetic energy loss. In the dilute limit ($x=0.0625$) for large enough
$\mathcal{D}$, we have that defect realizations with shared vertical
bonds become so much more favorable and, therefore, so much more easy
to converge numerically that they dominate the statistical averages.
Just for this doping, at $\sim\mathcal{D}_c$, one has a transition
from a situation with mainly well separated defect cubes to
mainly couples of defect cubes sharing one vertical bond, see
Fig. 5(d). As a matter of fact, $\mathcal{D}$ affects also the way
polarons interact: below $\mathcal{D}_{c}$, they avoid each other to 
maximize the gain in kinetic energy, but above $\mathcal{D}_{c}$, they 
attract each other in order to minimize the loss in kinetic energy.

Indeed, $\mathcal{D}_c$ corresponds to the value of $\mathcal{D}$ that 
imposes an orbital rotation large enough to make the kinetic energy 
gain on an active bond equal to just half of that of a standard 
superexchange bond, making thus equal the kinetic energy gain on two 
separated defect cubes (two \emph{active bonds} and six 
\emph{spectator bonds} \cite{Ave18}) and two defect cubes 
sharing an empty vertical bond: no active bonds, with no kinetic 
energy gain at all on the empty shared vertical bond, still six 
spectator bonds, but a whole ordinary superexchange bond
recovered! Such an occurrence makes clear why at $\mathcal{D}_{c}$
the kinetic energy per defect is almost completely independent of $x$ 
(we have an approximate isosbestic point). The value of $x$ just rules 
the number of defect cubes sharing an empty vertical bond and if 
the energy of the two relevant configurations (two active bonds 
or one empty vertical bond plus a recovered ordinary superexchange 
bond) is equal, the dependence on $x$ is clearly lost. The small loss 
of kinetic energy $\delta k$ at $\mathcal{D}_{c}$, is then equal to 
that of a missing ordinary superexchange bond plus the difference 
between the kinetic energy of six ordinary superexchange bonds and the 
kinetic energy of six rotated spectator bonds.

It is now clear that the presence of defect cubes sharing empty
vertical bonds (forming bipolarons) is very relevant --- for larger and 
larger values of doping this comes to dominate the physical properties. 
This is extremely clear by looking at $m_{a+b}^{o}$ as a function of 
$x$ for all finite values of $\mathcal{D}$. On increasing 
$x>x_{c}\approx0.3$, one adds defects in V cubes that have almost all 
corners/edges already belonging to other defect cubes \cite{Domb}. 
Hence, no substantial increase of rotations and decrease of orbital 
order above a certain value of $\mathcal{D}=0.07>\mathcal{D}_{c}$. 
Each added hole either generates a polaron (on an active bond) 
or a bipolaron (on an empty vertical bond). The latter actually 
slightly increases the orbital order, as it is clearly shown by  
$m_{a+b}^{o}$ versus $x$ in the regime of large values of 
$\mathcal{D}$. The position of the minimum defines $x_{c}$ where the 
balance is reached between decreasing the orbital order through the 
the formation of new spectator bonds and its increase by changing 
polarons into bipolarons. It also coincides with value at which $K$, 
as a function of $x$ for $\mathcal{D}=0$, passes through the undoped 
value. The gain of kinetic energy at polarons (active bonds) is again
balanced by its loss at bipolarons (empty bonds).

Just one final remark regarding the role of $\mathcal{D}_{c}$ in the
IPN: above $\mathcal{D}_{c}$, in the middle of the LHB, and less
evidently right below $\mu$, the delocalization inverts its overall
trend and actually decreases as the orbital polarization inhibits the 
gain of kinetic energy along the active bonds. Right above
$\mu$, this mechanism works for all values of $\mathcal{D}$.


\begin{thebibliography}{99}

\bibitem{Kei15} B. Keimer, S. A. Kivelson, M. R. Norman, S. Uchida, J. Zaanen,
                   Nature \textbf{518}, 179 (2015).

\bibitem{Ima98} M. Imada, A. Fujimori, and Y. Tokura,
                   Rev. Mod. Phys. \textbf{70}, 1039 (1998).

\bibitem{Gru18} F. Grusdt, M. K\'anasz-Nagy, A. Bohrdt, C. S. Chiu,
                   G. Ji, M.~Greiner, D. Greif, and E. Demler,
                   Phys. Rev. X  \textbf{8}, 011046 (2018);
                A. Bohrdt, D. Greif, E. Demler, M. Knap, and F. Grusdt,
                   Phys. Rev. B \textbf{97}, 125117 (2018).

\bibitem{Zho17} Y. Zhou, K. Kanoda, and T.-K. Ng,
                   Rev. Mod. Phys. \textbf{89}, 025003 (2017).

\bibitem{Uch91} S. Uchida, T. Ido, H. Takagi, T. Arima, Y. Tokura,
                   and S.~Tajima,
                   Phys. Rev. B \textbf{43}, 7942 (1991);
                B. Keimer, N.~Belk, R.~J.~Birgeneau, A. Cassanho,
                   C. Y. Chen, M.~Greven, M.~A.~Kastner, A. Aharony,
                   Y. Endoh, R. W. Erwin, and G.~Shirane,
                   \textit{ibid.} \textbf{46}, 14034 (1992).

\bibitem{Kas98} M. A. Kastner, R. J. Birgeneau, G. Shirane, and Y. Endoh,
                   Rev.~Mod. Phys. \textbf{70}, 897 (1998).

\bibitem{Lee06} P. A. Lee, N. Nagaosa, and X.-G. Wen,
                   Rev. Mod. Phys. \textbf{78}, 17 (2006).

\bibitem{Sca12} D. J. Scalapino,
                   Rev. Mod. Phys. \textbf{84}, 1383 (2012).

\bibitem{Tacon} G. Ghiringhelli, M. Le Tacon, M. Minola, S. Blanco-Canosa,
                   C. Mazzoli, N. B. Brookes, G. M. De Luca, A. Frano,
                   D. G. Hawthorn, F. He, T. Loew, N. Moretti Sala,
                   D. C. Peets, M. Salluzzo, E. Schierle, R. Sutarto,
                   G. A. Sawatzky, E. Weschke, B. Keimer, and L. Braicovich,
                   Science \textbf{337}, 6096 (2012).

\bibitem{Fra15} E. Fradkin, S. A. Kivelson, and J. M. Tranquada,
                   Rev. Mod. Phys. \textbf{87}, 457 (2015).

\bibitem{Tok06} Y. Tokura,
                   Rep. Prog. Phys. \textbf{69}, 797 (2006).

\bibitem{Dag01} E. Dagotto, T. Hotta, and A. Moreo,
                   Phys. Rep. \textbf{344} 1 (2001).

\bibitem{Qui98} M. Quijada, J. \v{C}erne, J. R. Simpson, H. D. Drew,
                   K. H. Ahn, A. J. Millis, R. Shreekala, R. Ramesh,
                   M. Rajeswari, and T.~Venkatesan,
                   Phys. Rev. B \textbf{58}, 16093 (1998).

\bibitem{Kil99} R. Kilian and G. Khaliullin,
                   Phys. Rev. B \textbf{60}, 13458 (1999).

\bibitem{Gio08} G. Giovannetti, S. Kumar, J. van den Brink, and S. Picozzi,
                   Phys. Rev. Lett. \textbf{103}, 037601 (2009).

\bibitem{Yun07} S. Yunoki, A. Moreo, E. Dagotto, S. Okamoto,
                   S. S. Kancharla, and A. Fujimori,
                   Phys. Rev. B \textbf{76}, 064532 (2007).

\bibitem{Cha12} A. Charnukha, A. Cvitkovic, T. Prokscha, D. Pr\"opper,
                   N.~Ocelic, A. Suter, Z. Salman, E. Morenzoni,
                   J. Deisenhofer, V.~Tsurkan, A. Loidl, B. Keimer,
                   and A. V. Boris,
                   Phys. Rev. Lett. \textbf{109}, 017003 (2012).

\bibitem{Cao16} Y. W. Cao, X. R. Liu, M. Kareev, D. Choudhury, S. Middey,
                   D. Meyers, J. W. Kim, P. J. Ryan, J. W. Freeland, and
                   J.~Chakhalian,
                   Nature Comm. \textbf{7}, 10418 (2016).

\bibitem{Kue17} L. Kuerten, C. Richter, N. Mohanta, T. Kopp, A. Kampf,
                   J. Mannhart, and H. Boschker,
                   Phys. Rev. B \textbf{96}, 014513 (2017).

\bibitem{Lei14} H. Lei, W.-G. Yin, Z. Zhong, and H. Hosono,
                   Phys. Rev. B \textbf{89}, 020409(R) (2014).

\bibitem{Kha01} G. Khaliullin, P.~Horsch, and A. M. Ole\'{s},
                   Phys. Rev. Lett. \textbf{86}, 3879 (2001);
                   Phys. Rev. B   \textbf{70}, 195103 (2004).

\bibitem{Kha05} G. Khaliullin,
                   Prog. Theor. Phys. Suppl. \textbf{160}, 155 (2005).

\bibitem{Ray07} M. De Raychaudhury, E. Pavarini, and O. K. Andersen,
                   Phys. Rev. Lett. \textbf{99}, 126402 (2007).

\bibitem{Yan04} J.-Q. Yan, J.-S. Zhou, and J. B. Goodenough,
                   Phys. Rev. Lett. \textbf{93}, 235901 (2004);
                J.-S. Zhou, J. B. Goodenough, J.-Q. Yan, and Y. Ren,
                   \textit{ibid.} \textbf{99}, 156401 (2007);
                J.-Q. Yan, J.-S. Zhou, J.~B.~Goodenough, Y. Ren, J. G. Cheng,
                   S. Chang, J. Zarestky, O. Garlea, A. Llobet, H. D. Zhou,
                   Y. Sui, W. H. Su, and R.~J.~McQueeney,
                   \textit{ibid.} \textbf{99}, 197201 (2007).

\bibitem{Reu12} J. Reul, A. A. Nugroho, T. T. M. Palstra,
                   and M. Gr\"uninger,
                   Phys. Rev. B \textbf{86}, 125128 (2012).

\bibitem{Kas93} M. Kasuya, Y. Tokura, T. Arima, H. Eisaki, and S. Uchida,
                   Phys. Rev. B \textbf{47}, 6197 (1993).

\bibitem{Tok00} S. Miyasaka, T. Okuda, and Y. Tokura,
                   Phys. Rev. Lett. \textbf{85}, 5388 (2000).

\bibitem{Fuj05} J. Fujioka, S. Miyasaka, and Y. Tokura,
                   Pnys. Rev. B \textbf{72}, 024460 (2005).

\bibitem{Fuj08} J. Fujioka, S. Miyasaka, and Y. Tokura,
                   Phys. Rev. Lett. \textbf{97}, 196401 (2006);
                   Phys. Rev. B     \textbf{77}, 144402 (2008).

\bibitem{Ave15} A. Avella, A. M. Ole\'{s}, and P. Horsch,
                   Phys. Rev. Lett. \textbf{115}, 206403 (2015).

\bibitem{Ave18} A. Avella, A. M. Ole\'{s}, and P. Horsch,
                   Phys. Rev. B \textbf{97}, 155104 (2018).

\bibitem{Ren00} Y. Ren, T. T. M. Palstra, D. I. Khomskii, A. A. Nugroho,
                   A.~A.~Menovsky, and G. A. Sawatzky,
                   Phys. Rev. B \textbf{62}, 6577 (2000).

\bibitem{Miy06} S. Miyasaka, Y. Okimoto, M. Iwama, and Y. Tokura,
                   Phys. Rev. B \textbf{68}, 100406(R) (2003);
                S. Miyasaka, J. Fujioka, M. Iwama, Y. Okimoto,
                   and Y. Tokura,
                   \textit{ibid.}  \textbf{73}, 224436 (2006).

\bibitem{Sol06} I. V. Solovyev,
                   Phys. Rev. B \textbf{74}, 054412 (2006).

\bibitem{Fuj10} J. Fujioka, T. Yasue, S. Miyasaka, Y. Yamasaki, T.~Arima,
                   H.~Sagayama, T. Inami, K. Ishii, and Y. Tokura,
                   Pnys. Rev. B \textbf{82}, 144425 (2010).

\bibitem{Yan11} J.-Q. Yan, J.-S. Zhou, J. G. Cheng, J. B. Goodenough,
                   Y. Ren, A. Llobet, and R. J. McQueeney,
                   Phys. Rev. B \textbf{84}, 214405 (2011).

\bibitem{Hor08} P. Horsch, A. M. Ole\'s, L.F. Feiner, and G. Khaliullin,
                   Phys. Rev. Lett. \textbf{100}, 167205 (2008).

\bibitem{Nog00} M. Noguchi, A. Nakazawa, S. Oka, T. Arima,
                   Y. Wakabayashi, H. Nakao, and Y. Murakami,
                   Phys. Rev. B \textbf{62}, R9271 (2000).

\bibitem{Bla01} G. R. Blake, T. T. M. Palstra, Y. Ren, A. A. Nugroho,
                   and A.~A.~Menovsky,
                   Phys. Rev. Lett. \textbf{87}, 245501 (2001);
                   Phys. Rev. B     \textbf{65}, 174112 (2002).

\bibitem{Sag08} M. H. Sage, G. R. Blake, and T. T. M. Palstra,
                   Phys. Rev. B \textbf{77}, 155121 (2008).

\bibitem{Ree16} M. Reehuis, C. Ulrich, P. M. Abdala, P. Pattison,
                   G. Khaliullin, J. Fujioka, S. Miyasaka, Y. Tokura,
                   and B. Keimer,
                   Phys. Rev. B \textbf{94}, 104436 (2016).

\bibitem{Hor11} P. Horsch and A. M. Ole\'{s},
                   Phys. Rev. B \textbf{84}, 064429 (2011).

\bibitem{Ave13} A. Avella, P. Horsch, and A. M. Ole\'{s},
                   Phys. Rev. B \textbf{87}, 045132 (2013).

\bibitem{Ole07} A. M. Ole\'{s}, P.~Horsch, and G. Khaliullin, 
                   Phys. Rev. B \textbf{75}, 184430 (2007).

\bibitem{Dag11} M. Daghofer, A. Nicholson, A. Moreo, and E. Dagotto,
                   Phys. Rev. B \textbf{81}, 014511 (2010);
                A. Georges, L. d'Medici, and J.~Mravlje,
                   Annu. Rev. Condens. Matter Phys. \textbf{4}, 137 (2013);
                K. M. Stadler, Z. P. Yin, J. von Delft, G. Kotliar,
                   and A. Weichselbaum,
                   Phys. Rev. Lett. \textbf{115}, 136401 (2015).

\bibitem{Ish05} S. Ishihara,
                   Phys. Rev. Lett. \textbf{94}, 156408 (2005).

\bibitem{Dag08} M. Daghofer, K. Wohlfeld, A. M. Ole\'s, E. Arrigoni,
                   and P.~Horsch,
                   Phys. Rev. Lett. \textbf{100}, 066403 (2008);
                P.~Wr\'obel and A.~M.~Ole\'s,
                   \textit{ibid.} \textbf{104}, 206401 (2010).

\bibitem{Bis15} V. Bisogni, K. Wohlfeld, S. Nishimoto, C. Monney,
                   Jan Trinckauf, K. Zhou, R. Kraus, K. Koepernik,
                   C. Sekar, V. Strocov, B.~B\"uchner, T. Schmitt,
                   J. van den Brink, and J. Geck,
                   Phys. Rev. Lett. \textbf{114}, 096402 (2015).

\bibitem{Bie16} K. Bieniasz, M. Berciu, M. Daghofer, and A. M. Ole\'s,
                   Phys. Rev. B \textbf{94}, 085117 (2016).

\bibitem{Yam18} M. G. Yamada, M. Oshikawa, and G. Jackeli,
                   Phys. Rev. Lett. \textbf{121}, 097201 (2018).

\bibitem{Kha00} G. Khaliullin and S. Maekawa,
                   Phys. Rev. Lett.  \textbf{85}, 3950 (2000).

\bibitem{Ole83} A. M. Ole\'s,
                   Phys. Rev. B \textbf{28}, 327 (1983).

\bibitem{Ant12} A. E. Antipov, I. S. Krivenko, V. I. Anisimov,
                   A. I. Lichtenstein, and A. N. Rubtsov,
                   Phys. Rev. B \textbf{86}, 155107 (2012).

\bibitem{suppl} For more details see in the Appendix--Supplemental Material.

\bibitem{Miz95} T. Mizokawa and A. Fujimori,
                   Phys. Rev. B \textbf{51}, 12880(R) (1995);
                                \textbf{54},  5368    (1996);
                                \textbf{56},  R493    (1997).

\bibitem{Miz99} T. Mizokawa, D. I. Khomskii, and G. A. Sawatzky,
                   Phys. Rev. B \textbf{60},  7309 (1999);
                                \textbf{61}, R3776 (2000);
                                \textbf{61}, 11263 (2000).

\bibitem{Miz01} T. Mizokawa, L. H. Tjeng, G. A. Sawatzky, G. Ghiringhelli,
                   O.~Tjernberg, N. B. Brookes, H. Fukazawa, S. Nakatsuji,
                   and Y.~Maeno,
                   Phys. Rev. Latt. \textbf{97}, 077202 (2001).

\bibitem{Noh05} H.-J. Noh, S.-J. Oh, B.-G. Park, J.-H. Park, J.-Y. Kim,
                   H.-D. Kim, T. Mizokawa, L. H. Tjeng, H.-J. Lin,
                   C. T. Chen, S.~Schuppler, S. Nakatsuji, H. Fukazawa,
                   and Y.~Maeno,
                   Phys. Rev. B \textbf{72}, 052411 (2005).
                   
\bibitem{note1} See discussion of Fig. 20(d) in Ref. \cite{Ave13}.

\bibitem{Ole05} A. M. Ole\'s, G. Khaliullin, P. Horsch, and L. F. Feiner,
                   Phys. Rev. B \textbf{72}, 214431 (2005).

\bibitem{Kha93} G. Khaliullin and P. Horsch,
                   Phys. Rev. B \textbf{47}, 463 (1993).

\bibitem{Sah17} R. Saha, F. Fauth, V. Caignaert, and A. Sundaresan,
                   Phys. Rev. B \textbf{95}, 184107 (2017).

\bibitem{Domb}  C. Domb and N. W. Dalton, 
                   Proc. Phys. Soc. \textbf{89}, 859 (1966);
                \L{}.~Kurzawski and K. Malarz,
                   Rep. Math. Phys. \textbf{70}, 163 (2012).

\end{thebibliography}
\end{document}